# The simplest method to measure the geocentric lunar distance: a case of citizen science

Jorge I. Zuluaga[1,a], Juan C. Figueroa[2,b] and Ignacio Ferrin[1,c]

[1] *FACom - Instituto de Física - FCEN, Universidad de Antioquia, Calle 70 No. 52-21, Medellín, Colombia*

[2] *Independent Software Developer*

ABSTRACT

We present the results of measuring the geocentric lunar distance using what we propose is the simplest method to achieve a precise result. Although lunar distance has been systematically measured to a precision of few millimeters using powerful lasers and retroreflectors installed on the moon by the Apollo missions, the method devised and applied here can be readily used by nonscientist citizens (e.g. amateur astronomers or students) and it requires only a good digital camera. After launching a citizen science project called the "Aristarchus Campaign", intended to involve astronomy enthusiasts in scientific measurement of the Lunar Eclipse of 15 April 2014, we compiled and measured a series of pictures obtained by one of us (J.C. Figueroa). These measurements allowed us to estimate the lunar distance to a precision of ~3%. We describe here how to perform the measurements and the method to calculate from them the geocentric lunar distance using only the pictures time stamps and a precise measurement of the instantaneous lunar apparent diameter. Our aim here is not to provide any improved measurement of a well-known astronomical quantity, but rather to demonstrate how the public could be engaged in scientific endeavors and how using simple instrumentation and readily available technological devices such as smartphones and digital cameras, any person can measure the local Universe as ancient astronomers did.

## 1. Introduction

Ancient astronomers devised clever methods to measure the local Universe. Although to a scale many orders of magnitude larger than the size of any human being, diameters and distances to the Sun and the Moon and the diameter of Earth itself, can be measured using what in principle seem to be very simple astronomical methods. One of the best known methods, were originally proposed by Aristarchus of Samos (c. 310 – c. 320 BCE). He devised a method using lunar phases and solar and lunar eclipses for estimations assuming circular orbits, of the relative sizes and distances of the Sun and

a jorge.zuluaga@udea.edu.co
b jcfigueroa@gmail.com
c respuestas2013@gmail.com

the Moon[1,2]. Also well known are the methods proposed and applied by Eratosthenes of Cyrene (c. 276 B.C. - c. 195/194 B.C.) to demonstrate the sphericity of the Earth and the size of our planet[3].

Almost two millennia after these first attempts, astronomers have measured such quantities to exquisite precision. The radius of our planet for instance has been measured to a precision of millimeters even although its shape is very irregular[4]. The distance to the Moon has been measured to a precision of centimeters using powerful lasers and retro-reflecting arrays of mirrors placed on the Lunar surface by Apollo Astronauts[5,6]. The size and distance of the Sun have been determined by different astronomical methods including Venus and Mercury transits and more recently advanced techniques of radar ranging and celestial mechanics[7,8,9] As a result, the radius of the Sun and its distance to the Earth are presently known to a precision ranging from few kilometers to few meters respectively, even although they are 6 to 11 orders of magnitude larger than their present uncertainties.

As opposed to the ancient methods, most of these modern techniques are far from the realm and technological capacities of amateur astronomers and nonscientist citizens. As a result, the measurement of the local Universe seems to most of us a matter of sophisticated instrumentation and advanced scientific capabilities. This is contrary to the spirit of the relatively simple measurements proposed by Aristarchus, Eratosthenes, and other astronomers many centuries ago.

Ancient methods used to measure the local Universe are common place in Astronomy and Physics textbooks[10]. Moreover, many efforts to use them as educational tools in the classroom have also been devised and published[11-16]. Today t plethora of advanced and accessible technological devices such as smartphones, tablets, digital cameras and precise clocks, is opening a new door to the realm of "do-it-yourself-science" and from there to the possibility of measuring the local Universe by oneself. Almost all these devices come with a GPS receiver, fast internet connections and precise clocks synchronized to international time systems. This technological state-of-the-art is giving us incredible opportunities to involve science enthusiasts and in general, nonscientist citizens, in global scientific projects, where their role is to provide in situ measurements that will be otherwise prohibitively expensive for a professional scientific team. We called this scientific and sociological phenomenon, "Citizen Science"[17]. A recent example of the power of these "crowd-science" is the thousand of pictures and videos of the Chelyabinsk Impact Event that allowed several scientific groups around the world to study the phenomenon[18-21]. Other well-known project is "Globe-at-Night" project[22], a successful initiative intended to measure light pollution all around the world.

The recent Lunar Eclipse of 15 April 2014 was our own opportunity to launch a citizen science project. We called it the "Aristarchus Campaign"[23]. The original aim of the campaign was to make the experience of the eclipse not only an enjoyable personal experience, but also to provide an opportunity to perform simple measurements able to provide anyone with first order estimations of the size of the local Universe especially using modern technological devices. Beside just taking beautiful pictures of the phenomenon, we invited people to perform scientific observations, including taking normal pictures, but with proper scientific rigor. Our hypothesis was that such a simple experiment could make the experience of an otherwise common astronomical phenomenon, into a personal project to measure the Universe.

We proposed people to take one of 5 measurements that they could perform using very simple instruments. The first experiment was to take pictures of the full moon from rising to culmination. It required only a good digital camera mounted on a tripod. This is the experiment whose results are reported here. The second experiment was to take pictures of the moon before it entered into the penumbra and after it was completely eclipsed. The third experiment was to measure, as precise as possible, the time of the eclipse contacts. The fourth was to take good pictures of the partial eclipse when the shape and size of the Earth shadow is clearly visible. And the last and fifth experiment was to take a picture of the Moon including in the same field of view, Spica and Mars. Regretfully, the weather in Colombia, where we originally launched the campaign, made it impossible for many enthusiasts to perform most of the measurements, especially during the most interesting phases of the eclipse. In some places, people performed some of the proposed experiments, but only during the initial phases of the partial eclipse. We expect to repeat the same social experiment during the second lunar eclipse of this Tetrad, on 8 October 2014.

Here we report the results of the analysis of a series of pictures taken by one of us (J.C.) from El Retiro (Antioquia), a small town 50 kilometers away from Medellin (Colombia). An analysis of his pictures, by application of simple plane and spherical trigonometry as well as an easy-to-program statistical procedure, allowed us to estimate the Earth-Moon distance to first order.

Two other papers have been recently published[15,16] showing how to measure the lunar distance and other properties of the lunar and Earth orbit using the same phenomenon we exploit here. One of them[15] performed careful measurements of the moon's apparent diameter at different zenithal angles using semi-professional astronomical equipment. The other one[16], performed more simple measurements using a hand-held camera. The results of the latter are far less precise than that of the former, but still, their instruments and methods are closer to our goal here: to make the procedure as simple as possible in order to be applied by a citizen scientist. Our method improves on those other attempts by minimizing the number of measured variable. We require that along with pictures, the time elapsed between the beginning of the experiment to the moment at which each picture is taken be measured to a precision of seconds. Since precise clocks are already incorporated in modern digital cameras, our method only requires the pictures and the timestamps that the camera puts on them. The elevation and the zenith angle of the moon at the moment when each picture is taken is required, a common condition in the methods published previously works. Although our "minimalistic method" is not as precise as the methods and results reported in[15], they are precise enough for the level required in a "citizen-science" project.

This report is organized as follows: in Section 2 we describe the method devised here to measure Earth-Moon distance using just a digital camera. Section 3 presents the measurements performed during the night of the 15 April 2014 Lunar Eclipse in El Retiro (Antioquia, lat. 6°7'40.86"N, long. 75°31'45.78"W). In Section 4, we present the results of the analysis of the pictures and the measured Earth-Moon distance. Section 5 is devoted to discuss the results and the limitations and potential improvements of the method. Finally Section 6 present the conclusions of what we consider the first report of the recently launched "Aristarchus Campaign".

## 2. Method

Although our perception seems to indicate that the Moon is closer (larger) when it is just above the horizon than when it is high in the sky[24], it is actually the opposite. The distance from the moon to any observer in the surface of the Earth, decreases as the moon rises in the sky, even although the moon distance to the center of the Earth remains approximately constant during the same time. In Figure 1 we show the cause of this nightly observer-moon distance variation.

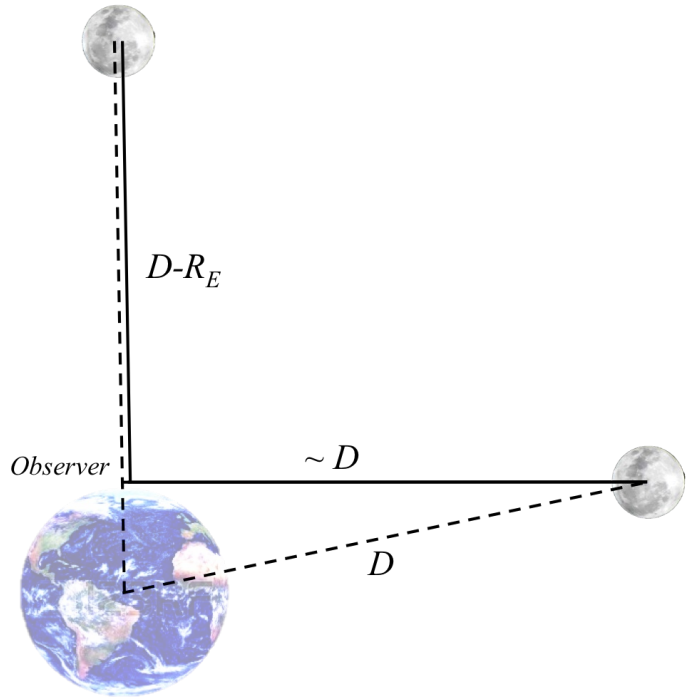

*Figure 1. When the moon rises its distance to an observer in the surface of the Earth is reduced. Objects are not shown to scale.*

The direct consequence of this variation is that the angular size of the moon is larger at culmination than in the horizon. In idealized conditions, the fractional change in angular diameter is as large as $\Delta\phi/\phi \sim R_E/D \sim 1/60 \sim 1.7\%$, where $\phi$ is the angular size of the moon, $R_E$ is the Earth Radius and $D$ is the geocentric lunar distance. This tiny apparent size variation is much lower than our naked eye angular resolution, but more importantly is far less than the variation that our brain perceives when the moon rises up from the horizon[24]. As a result, the phenomenon remained unperceived through centuries and, although relatively obvious, it has been scarcely exploited as a method to recognize the sphericity of the Earth or to measure the Moon distance relative to the Earth size in educative or outreach contexts.

Modern personal cameras have now reached the resolution level capable of capturing the moon with a precision enough to perceive and more importantly to measure this tiny variation in apparent size. If we assume that in a given optical configuration the Moon is able to occupy most of an electronic sensor having $p \times p$ pixels, an apparent size change of 1% is easily measurable if the number of pixels $p$ is larger than a few thousands, i.e. if the number of pixels in the Camera is larger than a few millions (Mpx). Many other variables such as the optical capabilities of the camera (in order to fill the sensor with the moon's image the camera should be able to make large optical zooms), the stability of the mounted tripod, vibrations induced by the wind, the stability of the atmosphere, among others, should also be taken into account. However a semi professional camera having 10X optical zoom or a telescopic objective, a sensor of ~10 Mpx and mounted on a solid tripod, could be enough to register the angular change of the moon during the night. Although "semi professional" sounds like an expensive and inaccessible equipment most of commercial digital cameras are starting to fell in this category.

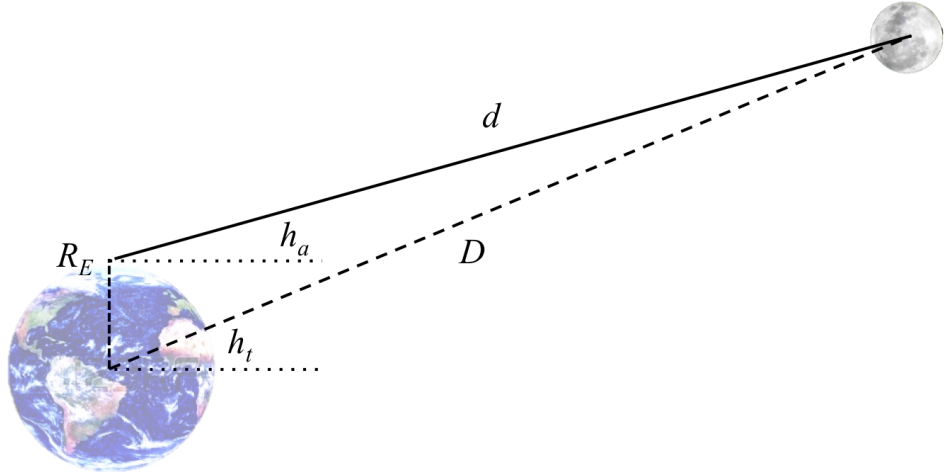

*Figure 2. The instantaneous distance of the moon (d) changes as the moon rises and its true and apparent elevation increase. This happens even although the geocentric distance of the Moon is unchanged*

The instantaneous angular diameter of the moon $\varphi(t)$ (apparent size) is related to its radius $R_L$ and instantaneous distance to the observer $d(t)$ by:

$$\varphi(t) = \frac{2R_L}{d(t)} \qquad (1)$$

The distance $d$ of a point in the surface of the Earth (radius $R_E$) to the Moon located at a distance $D$ as measured from the center of the Earth (geocentric lunar distance) are related by the cosine rule through (see Figure 2),

$$d(t)^2 = D^2 + R_E^2 - 2 D R_E \sin h_t(t) \qquad (2)$$
$$D^2 = d(t)^2 + R_E^2 + 2 d(t) R_E \sin h_a(t) \qquad (3)$$

where $h_t(t)$ and $h_a(t)$ are the geocentric and topocentric (true and apparent) elevations over the horizon (see Figure 2). As we see, the apparent size depends on D in a non-trivial way and changes in time as the Earth rotates.

In idealized conditions, when $h_t = 0°$, the observer-moon distance is $d(rising) = D^2 + R_E^2$ which neglecting second order terms in $R_E/D$ is just d ≈ D. When $h_t = 90°$ the distance is $d(culmination) = D - R_E$, i.e. almost 6400 km smaller. The fractional change between both extremes is $\Delta d/d(rising) \approx R_E / D$.

If we were able to measure the angular diameter of the Moon, $\varphi_0$ and $\varphi$, at two different times during the night when the elevation over the horizon were $h_{t,0}$ and $h_t$, and the distance to the observer was $d_0$ and $d(t)$, Eqs. (1) and (2) can be combined to give us:

$$\frac{d(t)}{R_E} = \frac{2[\sin h_t(t) - \alpha(t)\sin h_{t,0}]}{\alpha(t)^2 - 1} \qquad (4)$$

where we have introduced the relative apparent size:

$$\alpha \equiv \frac{\varphi_t(t)}{\varphi_0} = \frac{d_0}{d(t)} \qquad (5)$$

Equation (4) is a rather simple formula providing us with an estimation of the instantaneous lunar distance as a function of time, provided we can measure the relative apparent size of the moon $\alpha$ and its true elevation over the horizon $h_t$.

At first order in $R_E/D$ we can use Eq. (4) and the apparent elevation of the moon $h_a$ to obtain the instantaneous lunar distance $d$. Moreover, putting this result into Eq. (3) we can estimate the geocentric lunar distance $D$, thus completing the stated goal.

## 2.1. Increasing the precision, simplifying the measurement procedure

Equation (4) requires at least two measurements: the apparent diameter of the moon (arbitrary units, e.g. in pixels) and its apparent elevation in the sky (or equivalently its zenith angle), both at two different times during the night. Apparent sizes can be obtained from pictures and the elevation could be measured using one of the many apps able to calculate the tilt of a smartphone or a tablet with respect to the horizon. The last quantity, however, could be affected by significant systematic errors, rendering the estimation of lunar distance very uncertain. This is true even if the apparent diameter is measured exquisitely well in pictures. For these reasons previous attempts to use this method, have normally required semi professional equipment[15] if one has any hope of obtaining high precision results[16].

In order to increase the precision of the estimated lunar distance and to avoid the measurement of the apparent elevation of the moon (which moreover is not exactly the geocentric elevation we need in Eq. (4)) we have devised a simple mathematical and statistical procedure. Although the procedure could seem beyond the skills of a normal "citizen scientist" it is actually very simple when compared with other astrometric or dynamical methods used in Astronomy.

The true elevation above the horizon $h_t$ of any celestial body increases in time due to the rotation of the Earth. Therefore it can be computed from the equatorial declination $\delta$ and hour angle $H(t)$ of the moon using the well-known spherical trigonometry relationship:

$$\sin h_t = \sin l \sin \delta + \cos l \cos \delta \cos H(t) \qquad (6)$$

here $l$ is the geographic latitude of the observing site that could be measured with high precision using the GPS receiver of any smartphone. The hour angle can be calculated from a precise measurement of the time elapsed from the first observation $(t-t_0)$ using:

$$H(t) = H_0 + H_r(t - t_0) \qquad (7)$$

where $H_0$ is the hour angle at a reference time $t_0$ and $H_r$ = 14.959043495°/h is the rate of celestial daily motion. Since the moon has a mean proper motion in right ascension of 0.549014937°/h (360°/27.321661 days) the effective rate of hour angle advancement will be $H_{r;Moon}$ = 14.41002856 °/h.

According to Equation (6) and (7) if we very precisely measure the time at which the pictures are taken and assume proper values for $H_0$ and $\delta$, we will be able to compute the lunar distance using the model presented in Section 2. However, the precision of the estimation will now be limited by the values of these two free parameters. One way to "get rid" of these unknown quantities is to perform more than two measurements during the night. Instead of taking a couple of pictures, one at Moon rise and one at culmination, we can take several pictures at very different elevations all night long. As a result, we will have a set of measurements of $(t-t_0)$ and $\phi$ that can be fitted with our model (Eqs. 2, 4-7) This fitting procedure provide us three important unknown quantities: $D$, $H_0$ and $\delta$.

We apply this procedure to a series of actual pictures taken during the night of 14 April 2014 and the morning of 15 April 2014 as part of the "Aristarchus Campaign".

## 3. Observations and Measurements

The weather conditions were very variable during the observation. The observation started with the East horizon covered by clouds making it impossible to take pictures of the Moon rise. The Moon was completely visible above the horizon approximately at 18:27:09 LMT and at this time started the photographic session. During the 7 hours of observation, the weather conditions changed constantly. The last picture was taken at 24:59:14, one hour after the lunar culmination. In total, we took 20 pictures. Due the cloud coverage, exposure time of the photos varies between 1/2 and 1/100 of second, preserving constant the sensitivity ISO 100.

The photos were taken with an entry-level DSLR Canon Rebel T3i (600D), with a 22.3 x 14.9 mm CMOS sensor and a resolution of 5184px x 3456px (18 MP). The camera clock was synchronized with the UTC time one hour before to start the observation and the photos were taken in RAW and JPEG to ensure maximum quality. The camera was attached using a T-Ring and T-Adapter to the primary focus of an amateur Celestron Powerseeker 60AZ, with 60mm of aperture, 700mm of focal length and an altazimuth mount.

Before each photo was taken, a precise focus was obtained by a straightforward technique: (1) Activating the live mode of the camera, (2) pointing the telescope to some edge of the Moon, (3) applying the 10x digital zoom and (4) moving the focus knob of the telescope while observing the details of the craters in order to determine the precise focus to use.

Among the pictures we selected after a simple visual inspection a total of 12 pictures having the most suitable properties for the purposes of the Campaign. In Table 1 we summarize the timing and properties of the pictures used in our analysis.

Beside pictures we perform three additional measurements: (1) the precise location of the observing site, (2) the approximate elevation of the Moon at the time of the first picture and (3) the azimuth of the moon at this same moment. The geographic position was measured using the application "GPS Status"

v4.4.86 running on an Android v4.2.1 cellphone. The cellphone was located at a fixed place for 5 minutes and according to the application, the position was obtained after triangulating with 10 satellites. We assume that the precision of the 3 dimensional position is of the order of a few meters, or equivalently $10^{-5}$ degrees in latitude. The elevation of the Moon at the time of the first picture was estimated using the app "Camera Angle" v1.0.2. For that purpose we placed the cellphone over the telescope tube and read the elevation in the screen. The value obtained was 11º. We estimate an uncertainty of 2º for this measurement. Finally the azimuth of the Moon was measured using the digital compass of the cellphone. Independently, we used an analog compass and corrected for magnetic declination. Both measurements were consistent with an azimuth of 101º at the time of the first picture. An uncertainty of 2º was also assumed for this measurement, based on consistency of the methods.

*Table 1. Summary of the measured quantities.*

| Picture Id. | Time | Apparent Diameter (px) | Error (px) |
|---|---|---|---|
| 3105 | 18:30:37 | 746.29 | 3.92 |
| 3107 | 21:25:38 | 755.51 | 1.27 |
| 3108 | 21:29:11 | 755.21 | 2.72 |
| 3109 | 21:31:54 | 754.37 | 1.20 |
| 3110 | 21:35:49 | 754.30 | 1.93 |
| 3113 | 21:40:11 | 754.61 | 1.55 |
| 3134 | 23:45:14 | 757.14 | 1.20 |
| 3135 | 23:53:37 | 758.35 | 1.01 |
| 3136 | 23:57:30 | 756.29 | 1.19 |
| 3137 | 24:09:22 | 758.30 | 1.19 |
| 3138 | 24:11:21 | 756.92 | 1.08 |
| 3139 | 24:12:09 | 758.52 | 1.49 |

Figure 3 shows 6 of the 12 the pictures taken during the 14 April night and 15 April morning. A visual inspection of the images shows how hard is for the human eye to notice the subtle change of the apparent size of the moon from rising (first picture to the left) to culmination (last picture to the right).

## 4. Results

### 4.1. Moon apparent size

In order to perform the analysis described in Section 2 we need first to measure the Moon's apparent diameter directly from the pictures. Although at first sight it seems a rather simple procedure that can be performed with any commercial or free image processing software, reaching the desired level of precision implies a considerable effort. After trying several procedures we developed an algorithm able to provide not only reliable values of the Moon's diameter but also the minimum uncertainties in the measured quantity.

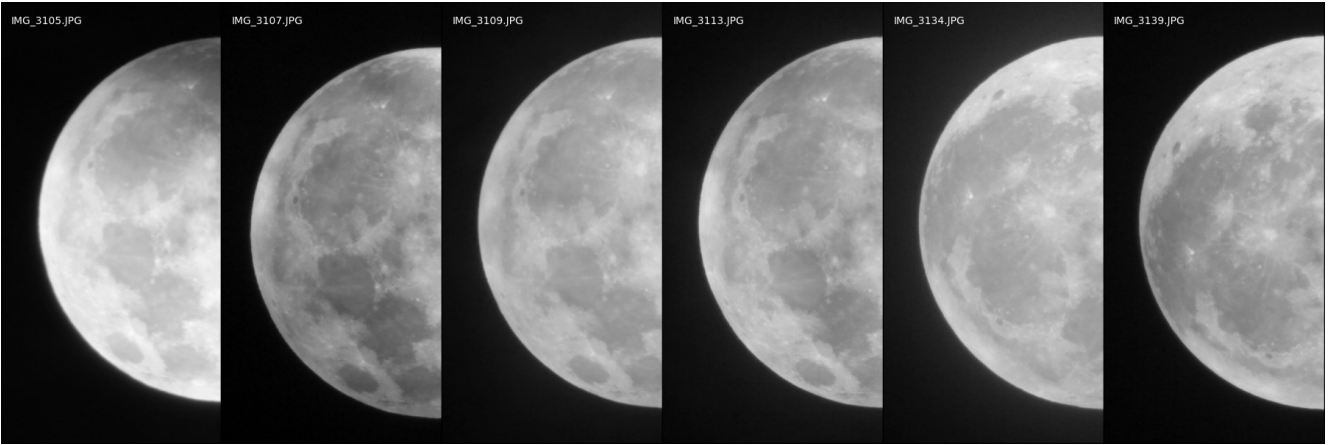

*Figure 3. Half-moons of 6 of the pictures used in this work. The change in apparent size is subtle and undetectable to visual inspection.*

In order to detect the Moon's edge we first apply a filter to the image, such that pixels having a value below a given threshold *T* received a new value of 0. Conversely, pixels above *T* received a value of 1. Once downgraded to this black and white scale, we apply an automatic contour procedure included in the `Python Matplotlib`[25] Library. This procedure attempts to find the coordinates *(x,y)* over the image of the pixels having a constant value *L*. We found after repeated experiments that a value *L*=0.5 provides the most reliable results for our purpose.

In idealized conditions the contour provided by the previous procedure will be the edge of the moon. Therefore if we use the coordinates of the contour and average the values of *x* and *y* we can estimate the position of the lunar image centroid. The Moon's radius in the image, *r*, is calculated as the arithmetic average of the distance from each contour point to the centroid. The uncertainty in the radius, *Δr* is assumed simply to be the dispersion of the contour-centroid distances.

The results of the procedure will depend on the arbitrary value of the threshold *T* used to filter the image in the first step. In Figure 4, we show the contours obtained using different values of *T* over the same image. As illustrated, low and high values of *T* are not suitable to detect the Moon's edge. On one hand, a low *T* detects the noise in the black sky far from the moon. On the other hand, a large *T* produces contours that delineate the lunar maria. However, there is an intermediate value of the threshold, $T_{opt}$, at which the contour delineates the lunar border almost perfectly. In our algorithm $T_{opt}$ was found by determining the threshold at which the dispersion *Δr* was minimum. This working definition is in agreement with the idea that the Moon's border, being an actual physical edge, will have the minimum dispersion, as opposed to other borders in the image. Finally the apparent size of the moon and its uncertainty, were taken to be the values *r* and *Δr* as obtained with the optimal threshold $T_{opt}$.

We have carefully verified the results of this automatic procedure comparing the apparent sizes with those obtained after an independent manual determination of moon radius using a commercial image processing software. The results are in agreement within the estimated uncertainties. In Table 1 we report the radius of the moon in pixels and their respective uncertainties as measured over the selected 12 images in this report. We see that our procedure is precise enough to return the moon apparent size to a precision of a few pixels as expected for a large solid object as the moon.

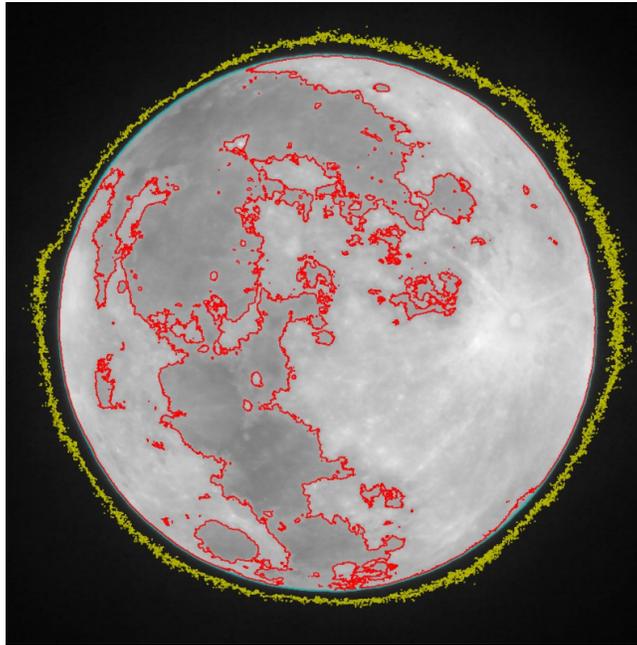

*Figure 4. Contours calculated on black and white version of the image obtained using different thresholds. The outer contour (yellow in the color figure) was obtained with a low threshold. Points of the noisy external areas of the moon image were detected. The inner contour (red) uses a large threshold. As a result the contour of the moon maria is detected. An intermediate value of the threshold (actually the optimum one, i.e. that providing the lowest dispersion in the radius) works very well at defining the Moon's border.*

**4.1. Fitting the moon distance**

Having the apparent sizes of the moon at different times during the night we can proceed to fit the theoretical model described in Section 2 in order to obtain our estimation of the geocentric lunar distance.

Our theoretical model has 4 free parameters: the geocentric lunar distance $D$, the initial hour angle $H_0$, the lunar declination $\delta$, and a geometric factor $f_0$ giving the expected size in pixels of the moon in the first picture. Four parameters give us too much room for fitting times and apparent sizes of the moon. Actually we have determined that a brute force fitting procedure provide us with a huge volume in the parameter space where the observations can be fitted. This is normal if we recognize that our method is trying to reduce the number of measured quantities as much as possible. In other methods the elevation of the moon at each time will be enough to constrain the model and reduce the size of the region in the parameter space where the solution lies. Our goal here, however, is to avoid introducing additional measurements, or better, to replace a hard to measure quantity such as the elevation with other much easier to measure such as the time. This replacement, however, has a cost.

To solve this statistical conundrum we performed a "constrained" fit, i.e. a fit where the free parameters are constrained to a priori known intervals. Constraining the interval of *D* was equivalent to cheating, so, we leave this parameter completely free. However the lunar declination and the initial hour angle are not as free as supposed. The hour angle of a rising moon should be larger than 18 hours and lower than 24 hours. On the other hand the declination of the Moon is never larger than 30º (north and south). Again, we discovered experimentally that these conservative constraints leave us too much room for a successful fit.

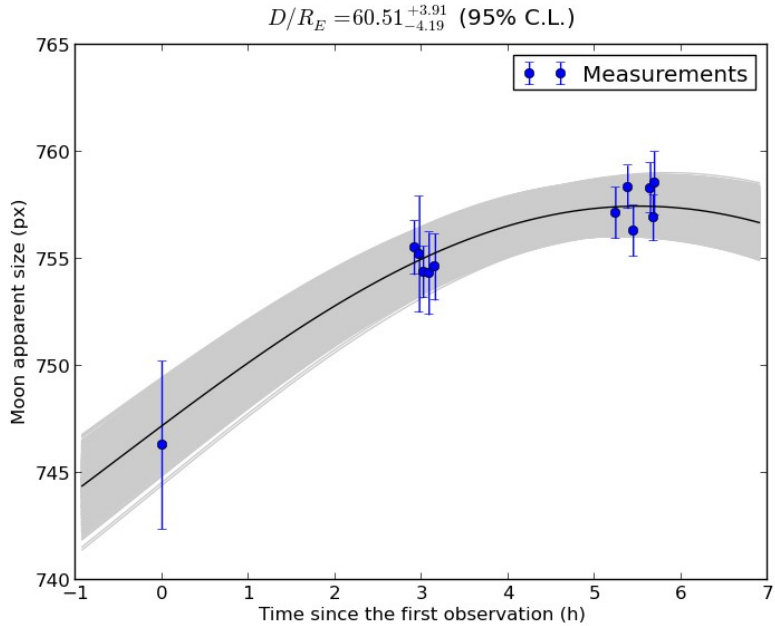

*Figure 5. Best fit of the measured apparent sizes (error-bars) to the theoretical model (continuous line). The shaded region correspond to solutions statistically compatible with the observed apparent sizes at a 5% confidence level.*

Our last resource was try to constrain these unknown quantities from the observed elevation and azimuth of the moon at the time of the first picture. We used this observation to show that the declination of the moon cannot be smaller than -15º or larger than -2º and that the hour angle was definitively between 18.2 hours and 19 hours.

Using these constrains we finally achieve consistent fits of the observed apparent diameters. However, our best fit analysis returned values of the free model parameters, too sensitive to the initial guesses. The main reason for this behavior was that the objective function (in this case the chi-square statistics) was too flat around the minimum value. To solve this problem we proceed by randomly exploring the four dimensional parameter space. We generate 300,000 4-tuples ($D, H_0, d, f_0$) around the best-fit values. Among these, we select those 4-tuples for which the chi-square of the observations given the model have a p-value larger than 0.05 and lower than 0.95 (95% two tailed C.L.) We found 1,484 sets of parameters satisfying this p-value criterium. These sets were the basis for our final estimation of the geocentric lunar distance and for its confidence level.

Figure 5 shows the result of applying this statistical procedure. The shaded region contains the theoretical curves of apparent size vs. time corresponding to the 1,484 sets of parameters satisfying the p-value criterium. The continuous line represents the model as evaluated by the mean of the free parameters of all the 1,484 sets.

This result implies that the geocentric lunar distance during the night of 14 April 2014 was $D/R_E = 60.51^{+3.91}_{-4.19}$ (95% C.L.) which is in excellent agreement with the theoretical average value for that night $D/R_E = 60.61$ as provided by the JPL Horizons Ephemeris System. In Table 2 we summarize the results of the statistical procedure.

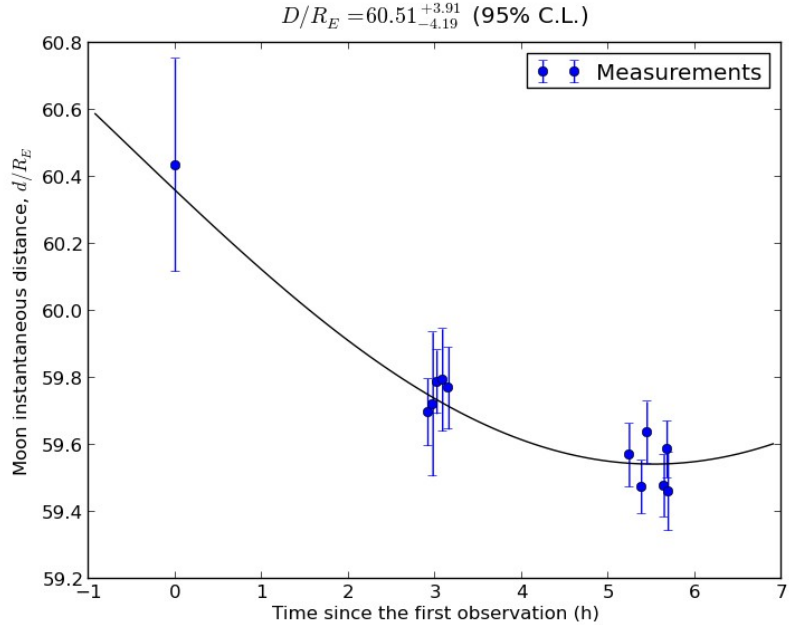

Figure 6. Instantaneous distance as a function of time elapsed since the first observation.

Table 2. Summary of the fitting results.

| Parameter | Monte Carlo Average and Dispersion | 95% C.L. | Actual Values |
| --- | --- | --- | --- |
| Geocentric Lunar distance, $D/R_E$ | 60.51 ± 2.32 | [56.32 , 64.41] | 60.53 – 60.69 $R_E$* |
| Lunar declination, $\delta$ | -9.54° ± 2.20 | [-13.27° , -5.69°] | -8.88° |
| Initial hour angle, $H_0$ | 18.67$^h$ | [18.35$^h$ , 18.98$^h$] | 18.71 h |

*During the 6 hour period of the observations the geocentric distance of the moon changed almost 1,000 kilometers.*

Finally, Figure 6 shows the value of the instantaneous lunar distance as a function of time as calculated from the observed apparent sizes and from the theoretical model as evaluated in mean value of the Monte Carlo sets of parameters.

## 5. Summary and discussion

We have devised a method to measure the geocentric lunar distance using only a set of high-resolution pictures taken during a night of full moon. The timestamps of the pictures and the apparent size of the moon in pixels provides the only additional information required to fit the lunar distance. In fact elevation or zenith angle measurements are required as assumed in previous works. Only a very crude estimation of the Moon's elevation and azimuth at the time of the first picture is actually required to constrain the fitting procedure and obtain a precise estimation of the lunar distance.

The uncertainty of our estimation here ($\Delta D$ = 2.32 $R_E$) is comparable to that obtained with a more careful procedure performed using semi professional equipment. This is actually very promising for a method relying only on the usage of a commercial digital camera and it shows the suitability of this method in Citizen Science initiatives.

How can this result be improved? In other words, which recommendation should others follow attempting the same measurements but looking for even lower uncertainties?

As observed in Figure 3, the critical factor determining the ability of our method to precisely estimate $D$, is the uncertainty in Moon's apparent size. But not the absolute uncertainty. Actually, the errors in our apparent size estimates were close to the theoretical limit of 1 pixel. It is the relative error, i.e. the absolute error divided by the observed size, that actually matters here. Working at the theoretical 1 pixel limit, relative errors could be reduced if we increase the size of the image over the sensor, for example through magnification.

A complete analysis of our observation shows us that the ideal conditions to perform the measurement combines several important factors such as (1) higher telescope apertures to improve the angular resolution of the borders of the Moon, (2) higher optical zoom to use more area of the camera sensor, (3) higher resolution cameras to have more pixels involved, (4) better weather conditions in order to reduce the seeing and finally (5) better techniques for focusing of automatized equipment.

Another critical factor controlling the precision of our estimation is a proper measurement of the Moon's size when close to the horizon. As observed in our own set of data the largest uncertainty in the apparent size of the moon corresponds precisely to the first image. A visual inspection of the image allows us to discover that the Moon's disk was too elongated for a full moon. This unusual eccentricity of the Moon's border was probably due to atmospheric refraction or other optical effects. Since our edge detection method is not sensitive to a direction dependent deformation of the image, we should pay special attention to how the algorithm tackles with images close to the horizon. On the other hand and if the relative uncertainties in apparent size estimations are reduced, one should avoid taking the first picture at very low elevation. This can be also accomplished by taking several pictures at the beginning when the Moon is rising.

The method to estimate the moon apparent size described in Section 4.1 works very well when the moon is completely full. A modification of the method to perform the same measurement when the moon is showing a different phase, should be pursued. It is important, however, to understand that any method devised for this purpose, should be able to provide realistic estimations of the apparent size errors. Any a priori estimation of these errors could make more complicated the statistical analysis.

## 6. Conclusions

We have measured the geocentric lunar distance applying what must be considered the be the simplest and affordable method yet devised. We determined that during the night of 14 April 2014 the Moon was at a relative distance $D/R_E = 60.51^{+3.91}_{-4.19}$ (95% C.L.). This value is in agreement with the true average value $D/R_E = 60.61$ as calculated for the same night with advanced dynamical models.

Our estimation resulted from the analysis of a series of pictures carefully taken by a citizen scientist, an initiative intended to involve enthusiasts and in general non-scientist citizens in scientific projects. In this particular case, the idea was to use the total lunar eclipse of 15 April 2014 to perform a series of measurements capable of providing us good estimations of the size of our local Universe.

We have demonstrated with this work not only that interesting astronomical results can be achieved using a very simple equipment and ingenious mathematical and statistical methods, but also that a there is a huge potential among non-scientists to contribute to scientific projects through careful measurements and detailed analyses.

**Appendix**

A copy of the original data and Python scripts developed for this work can be freely download from the research weblog of the FACom group at http://astronomia.udea.edu.co/aristarchus-campaign. There you will find two basic scripts:

- **get-moon-radius.py**: this script calculates the apparent radius of a *full moon* in a given image.

    - Usage:
        ```
        $ python get-moon-radius.py <IMAGE_FILE>
        ```
    - Returns:
        - Optimal threshold.
        - Apparent radius of the moon.
        - Minimum uncertainty in the apparent radius.
        - An image showing the moon and superimposed the contours at optimal and suboptimal thresholds.

- **fit-moon-size.py**: this script performs the statistical analysis of a set of apparent sizes measured over images of the full Moon.

    - Inputs: a plain data file with three columns: time of picture (HH:MM:SS), measured moon radius and radius uncertainty. An example input file is provided along with the scripts.

    - Configuration: before running the script you should configure several parameters:
        - OBSERVER_LATITUDE: latitude of the observing site.
        - MONTECARLO_POINTS: number of points used in the Monte Carlo sampling of the parameter space. Recommended 100,000.
        - H_INI, D_HINI: elevation of the moon (and its uncertainty) at the time of the first picture.
        - A_INI, D_AINI: azimuth of the moon (and its uncertainty) at the time of the first

picture.

- Usage:
  ```
  $ python fit-moon-size.py
  ```
- Returns:
  - 95% confidence level for the model parameters.
  - Plot of apparent sizes vs. time, including measurements and curves corresponding to the Monte Carlo consistent sets of model parameters.
  - Plot of instantaneous lunar distance vs. time, including measurements and curves corresponding to the Monte Carlo consistent sets of model parameters.

All the scripts require `Python >=2.7` and the libraries `Matplotlib`, `SciPy` and `NumPy`.

We will publish updates to the scripts and other interesting additional developments in the project weblog.

**Acknowledgements**

We want to thank to those friends and colleagues who support us through messages in social networks encouraging people to participate in the "Aristarchus Campaign". The Campaign is an initiative of the Sociedad Antioqueña de Astronomía (SAA) and University of Antioquia. We thank both organizations and institutions for promoting and supporting this kind of public initiatives. We also want to thank Paul Mason by the careful proofreading of the manuscript and insightful comments on its content.